\documentclass[aps,prb,floatfix,twocolumn,showpacs,reprint,superscriptaddress]{revtex4-1}
\usepackage{amsfonts}
\usepackage{mathrsfs}
\usepackage{amsmath}
\usepackage{color}
\usepackage{graphicx}
\usepackage{bm}
\usepackage{amssymb}
\usepackage{xspace}
\usepackage{epstopdf}
\usepackage{dcolumn}
\usepackage{longtable}
\usepackage{multirow}
\usepackage[colorlinks=true, letterpaper=true, pdfstartview=FitV, linkcolor=blue, citecolor=blue, urlcolor=blue]{hyperref}

\makeatletter

\newcommand{\Rmnum}[1]{\expandafter\@slowromancap\romannumeral #1@}
\makeatother

\begin{document}
\title{Buckled honeycomb lattice and unconventional magnetic response}

\author{Shengyuan A. Yang}
\affiliation{Engineering Product Development, Singapore University of Technology and Design, Singapore 138682, Singapore}

\author{Hui Pan}
\affiliation{Department of Physics, Beihang University, Beijing 100191, China}

\author{Fan Zhang}
\email{zhang@utdallas.edu}
\affiliation{Department of Physics, University of Texas at Dallas, Richardson, Texas 75080, USA}

\begin{abstract}
We study the magnetic response of buckled honeycomb-lattice materials. The buckling breaks the sublattice symmetry, enhances the spin-orbit coupling,  and allows the tuning of a topological quantum phase transition.
As a result, there are two doubly degenerate spin-valley coupled massive Dirac bands, which exhibit an unconventional Hall plateau sequence under strong magnetic fields.
We show how to externally control the splitting of anomalous zeroth Landau levels, the prominent Landau level crossing effects, and the polarizations of spin, valley, and sublattice degrees of freedom.
In particular, we reveal that in a p-n junction, spin-resolved fractionally quantized conductance appears in a two-terminal measurement with a spin-polarized current propagating along the interface.
In the low-field regime where the Landau quantization is not applicable, we provide a semiclassical description for the anomalous Hall transport.
We comment briefly on the effects of electron-electron interactions and Zeeman couplings to electron spins and to atomic orbitals.
\end{abstract}
\maketitle

\section{Introduction}
Two-dimensional (2D) atom-thin materials have been attracting significant attention in the past decades. Starting with the successful isolation of graphene sheets,\cite{novo2004,cast2009} several other 2D materials have been successfully synthesized or isolated, such as hexagonal boron nitride,\cite{wata2004} transition metal dichalcogenides,\cite{novo2005} silicene,\cite{lalm2010,vogt2012,fleu2012,chen2012} black phosphorus\cite{lli2014,xia2014,hliu2014} etc., and more such materials are theoretically predicted and await to be demonstrated experimentally.
With reduced dimensionality, these atom-thin materials exhibit distinct physical properties from three-dimensional materials and conventional 2D quantum well systems, and have triggered extensive studies aiming to utilize them for various applications.

Among these 2D materials, a class of materials have honeycomb-lattice geometries as their stable structures. For example, graphene, silicene,\cite{caha2009} the predicted germanene,\cite{caha2009} $X$-hydride/halide ($X$=N-Bi) monolayers,\cite{song2014,cliu2014} and stanene\cite{xu2013} all have such a kind of structure. As a result, they shared several common interesting properties in their electronic band structure. First, the low energy spectrum usually has two inequivalent valleys located at the hexagonal Brillouin zone corners known as $K$ and $K'$ points.
(Note that the last three materials have an extra valley at $\Gamma$ point, which we will comment in the Discussion section).
This valley degree of freedom has been proposed as a novel means to encode information,
and how to control and manipulate it have led to the concept of valleytronics.\cite{ryce2007}
Second, the honeycomb structure has two triangular sublattices, usually labeled as $A$ and $B$, which leads to a pseudospin structure of the electron wave function. When the sublattice (chiral) symmetry is broken, a band gap can be opened at K and K' points. This symmetry breaking could arise simply because two sites are occupied by different atoms, or as we are more interested in, because the lattice has buckling such that $A$ and $B$ sites have a relative shift along the direction perpendicular to the 2D plane.

In the presence of buckling, the inversion symmetry can be broken by a perpendicular electric field and the induced gap size can be tuned by controlling the field strength. In addition, the crystal symmetry allows the presence of an intrinsic spin-orbit coupling (SOC).\cite{kane2005a} This SOC is the key ingredient in the Kane-Mele quantum spin Hall (QSH) insulator originally proposed in graphene.\cite{kane2005a,kane2005b} Of course, the SOC strength is negligibly small in graphene.\cite{min2006,yao2007} Later, it was found that the SOC could be enhanced by the buckling due to the direct hybridization between $\pi$ and $\sigma$ orbitals, as being predicted for the case of silicene and germanene.\cite{liu2011a,liu2011b} Recently, several QSH insulators with large SOC gaps are proposed. In particular, theoretical analysis have revealed that for $X$-hydride/halide ($X$=N-Bi) monolayers,\cite{song2014} huge intrinsic SOC up to $1$ eV could arise because the low energy bands have $p_{x}$ and $p_y$ instead of $p_z$ orbital character (like graphene and silicene).\cite{song2014,cliu2014,cjwu2014} These distinct features lead to rich transport properties of these materials. Especially, the switch between the QSH and trivial insulating phases, tunable through an electric field,\cite{ezaw2012} may be utilized for designing energy efficient spintronic devices.

Although the Landau levels (LLs) of buckled honeycomb-lattice materials have been considered before,\cite{tabe2013a,tabe2013b,ezaw2012b,tahi2013}
a consistent and comprehensive analysis focusing on the transport regime is still lacking. In this paper, we will study the transport properties of this class of materials in response to an applied magnetic (orbital) field. As discussed above, the intrinsic buckling breaks the sublattice symmetry, enhances the spin-orbit coupling, and allows the tuning of inversion asymmetry. The resulting low energy spectrum thus splits into two sets of doubly degenerate spin-valley coupled massive Dirac fermions with different masses.
Importantly, the electric field is able to tune the mass difference and the quantum phase transition between the QSH and trivial insulating phases.
Under strong magnetic fields, the interplay between the SOC and the inversion asymmetry leads to an unconventional Hall plateau sequence.
Due to the mass difference, the LL spectrum shows prominent crossing effects.
Because the pseudospin chirality switches between the two valleys, the energies of the zeroth LLs are valley-dependent.
Hence it is possible to control the valley polarization of carriers by tuning doping level as well as external electrical and magnetic fields.
Moreover, as ideal candidates for bipolar nanoelectronics, these materials exhibit intriguing transport properties in a p-n junction geometry. Particularly, spin-resolved fractionally quantized conductance may appear in a two-terminal measurement, with a spin-polarized current propagating along the interface.
In the low-field or strong-disorder regime, although the Landau quantization may not be applicable, a semiclassical theory is possible to account for the anomalous Hall transport.

Our paper is organized as follows. In Sec.~II, we introduce the low energy effective model describing this class of buckled honeycomb-lattice materials, with emphases on the roles of the sublattice symmetry and the intrinsic buckling.
In Sec.~III, we derive the LL structures for the QSH and the trivial insulating phases, followed by discussions on the $SU(4)$ symmetry breaking of the anomalous zeroth LLs. We then analyze the LL crossing effects and the unconventional Hall plateau sequences for both phases in Sec.~IV.
In Sec.~V, we further study the two-terminal conductances in unipolar and bipolar regimes and find some extra integer and fractionally quantized plateaus.
In Sec.~VI, we reveal the possible electric-field control of the spin, valley, and sublattice polarizations.
We also provide a semiclassical theory in Sec.~VII for the anomalous Hall transport in the low-field regime where the Landau quantization is not applicable.
Finally, we conclude in Sec.~VIII with a discussion of the complexity added by the states at the $\Gamma$ point,
some speculations on the role of electron-electron interactions,
and an estimation of the Zeeman couplings to electron spins and atomic orbitals.

\section{Model Hamiltonian}
Materials with 2D honeycomb lattice structures such as graphene, silicene, germanene, and $X$-hydride/halide ($X$=N-Bi) monolayers have their low energy bands around two valleys at K and K' points of the hexagonal Brillouin zone. Without spin-orbit coupling, the energy gap of the two linearly dispersed bands closes at the Dirac points, i.e., K and K' points. Near each Dirac point the low energy Hamiltonian can be written as
\begin{equation}\label{h0}
H_0=\hbar v(\tau_z k_x\sigma_x+k_y\sigma_y),
\end{equation}
where $v$ is a material-specific Fermi velocity, $\tau_z=\pm 1$ labels the two valleys K and K', $\sigma$'s are Pauli matrices representing the AB sublattice degrees of freedom. It should be noted that for different materials, the low energy physics could be associated with different types of orbitals. For example, for graphene, silicene, and germanene, it is of $p_z$ orbital type,\cite{cast2009,liu2011a} whereas for $X$-hydride/halide ($X$=N-Bi) ($X$=N-Bi), it is of $p_x$ and $p_y$ orbital types.\cite{song2014,cliu2014,cjwu2014}

The gapless nature of Eq.~(\ref{h0}) is protected by the following sublattice (or chiral) symmetry
\begin{eqnarray}\label{SL}
\{H_0, \sigma_z\}=0
\end{eqnarray}
and the topological winding number $\pm 1$ of the constant energy contour of the positive or negative energy band at each valley. Thus, the Dirac fermions described by Eq.~(\ref{h0}) acquire an energy gap in the presence of sublattice symmetry breaking that violates Eq.~(\ref{SL}) or when the two valleys couple and annihilate with each other. At least three possible sources of sublattice symmetry breaking can arise in these Dirac materials. First, strong electron-electron interactions may lead to an antiferromagnetic order yielding a quasiparticle gap at the Dirac point. However, this mechanism is not likely for any material we discuss here as $v$ is not sufficiently small.~\cite{v-small}

\begin{figure}[t]
\scalebox{0.46}{\includegraphics*{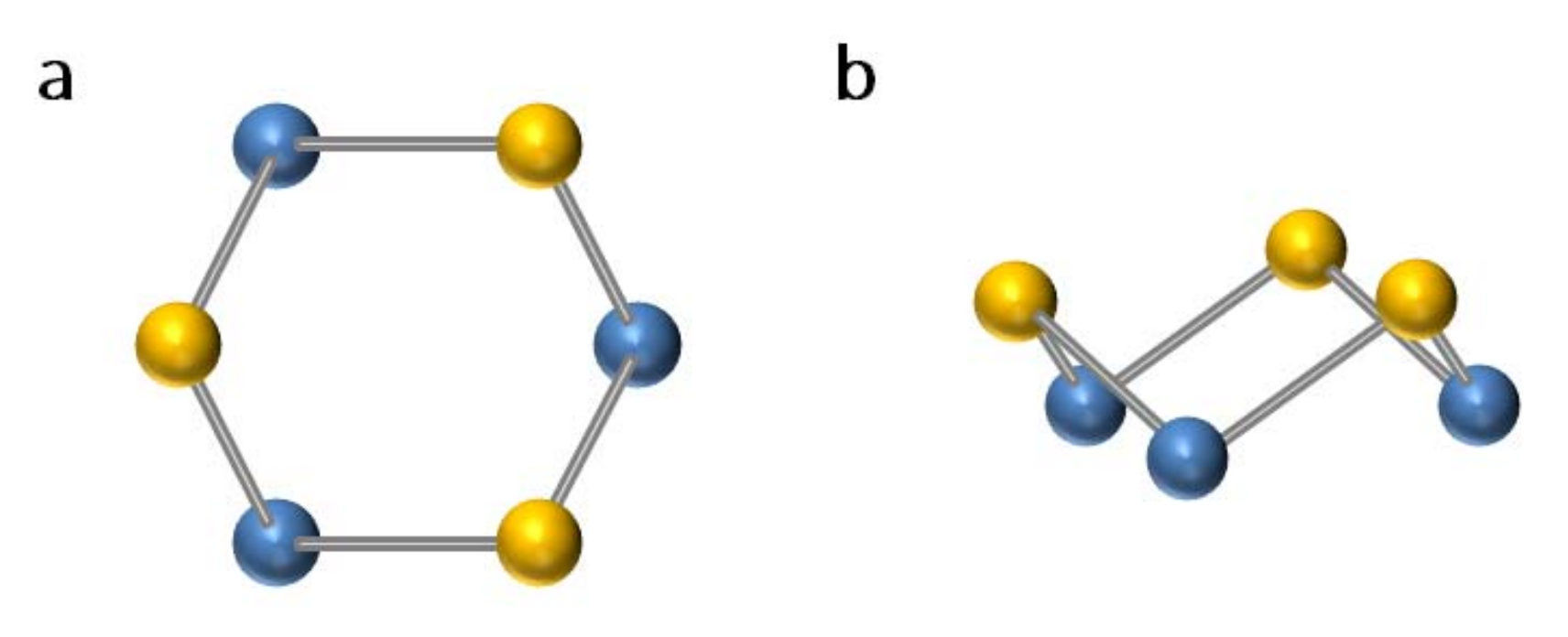}}
\caption{(color online) (a) Top view and (b) side view of the unit cell of a 2D buckled honeycomb lattice structure. The atomic sites of the two sublattices $A$ and $B$ are marked in different colors. The two sublattice planes have a relative shift along the perpendicular direction.}
\label{fig1}
\end{figure}

An energy difference between the two sublattices breaks inversion and sublattice symmetries producing a trivial band gap at the Dirac point, which can be modeled by
\begin{equation}\label{hg}
H_g=\lambda\sigma_z.
\end{equation}
Traditionally, this staggered sublattice potential $\lambda$ can hardly be induced in graphene. Recently, $\lambda$ can be even larger than $1$~eV when the two sublattices are formed by different atoms in hexagonal boron nitride and by different $d$ orbitals in MoS$_2$.\cite{xiao2012,xli2013} Of particular interest here, buckling of the two sublattices in opposite out-of-plane directions allows an easy way to control $\lambda$ via an external electric field perpendicular to the 2D plane (see Fig.~\ref{fig1}), although the buckling respects the inversion symmetry and does not lead to a nonzero $\lambda$. It has been found that such buckling of the honeycomb lattice is intrinsic for many monolayer materials including silicene, germanene, and $X$-hydride/halide ($X$=N-Bi).

Another essential ingredient in the low energy physics of these materials is the following intrinsic SOC that is allowed by the lattice symmetry,
\begin{equation}
H_{so}=\lambda_{so}\tau_z\sigma_z s_z\,.
\end{equation}
Here $s_z$ denotes the $z$-component of electron spin and $\lambda_{so}$ is the coupling strength. Evidently, this term breaks the spin $SU(2)$ rotational symmetry and violates the sublattice symmetry~(\ref{SL}), opening a band gap in Hamiltonian~(\ref{h0}). Interestingly, this term drives the system into a QSH insulator state protected by the time-reversal symmetry and/or the mirror symmetry.\cite{note-mirror} The pristine graphene is comprised of light carbon atoms forming a planar structure and $\lambda_{so}$ is negligibly small ($\sim 10^{-3}$ meV).\cite{min2006,yao2007} For silicene and germanene, buckling helps enhance the intrinsic SOC because of the hybridization between $\pi$ and $\sigma$ orbitals. Calculations based on a density functional theory have predicted that the SOC induced gap can reach $\sim 1.5$ meV for silicene and $\sim 23.9$ meV for germanene.\cite{liu2011a} Recently, a new class of bulked honeycomb-lattice materials $X$-hydride/halide ($X$=N-Bi) monolayers has been predicted.\cite{song2014} The low energy bands of these materials are of $p_x$ and $p_y$ orbital characters leading to an on-site SOC.\cite{song2014,cliu2014,cjwu2014} This fact, together with the buckling and the heavy $X$ atoms, substantially enhances the SOC gap to as large as $1$ eV (e.g., for BiH monolayer).

In the following, we shall focus on the generic model
\begin{equation}\label{Ht}
H=H_0+H_g+H_{so},
\end{equation}
describing a class of 2D materials with buckled honeycomb lattices at least including silicene, germanene, and
$X$-hydride/halide ($X$=N-Bi) monolayers. Other types of SOC terms, such as the Rashba-type SOC resulting from structural inversion symmetry breaking, may also exist, but their strengths are typically much smaller than the terms above hence are neglected in our treatment.\cite{liu2011b,ezaw2012}

\begin{figure}[t]
\scalebox{0.44}{\includegraphics*{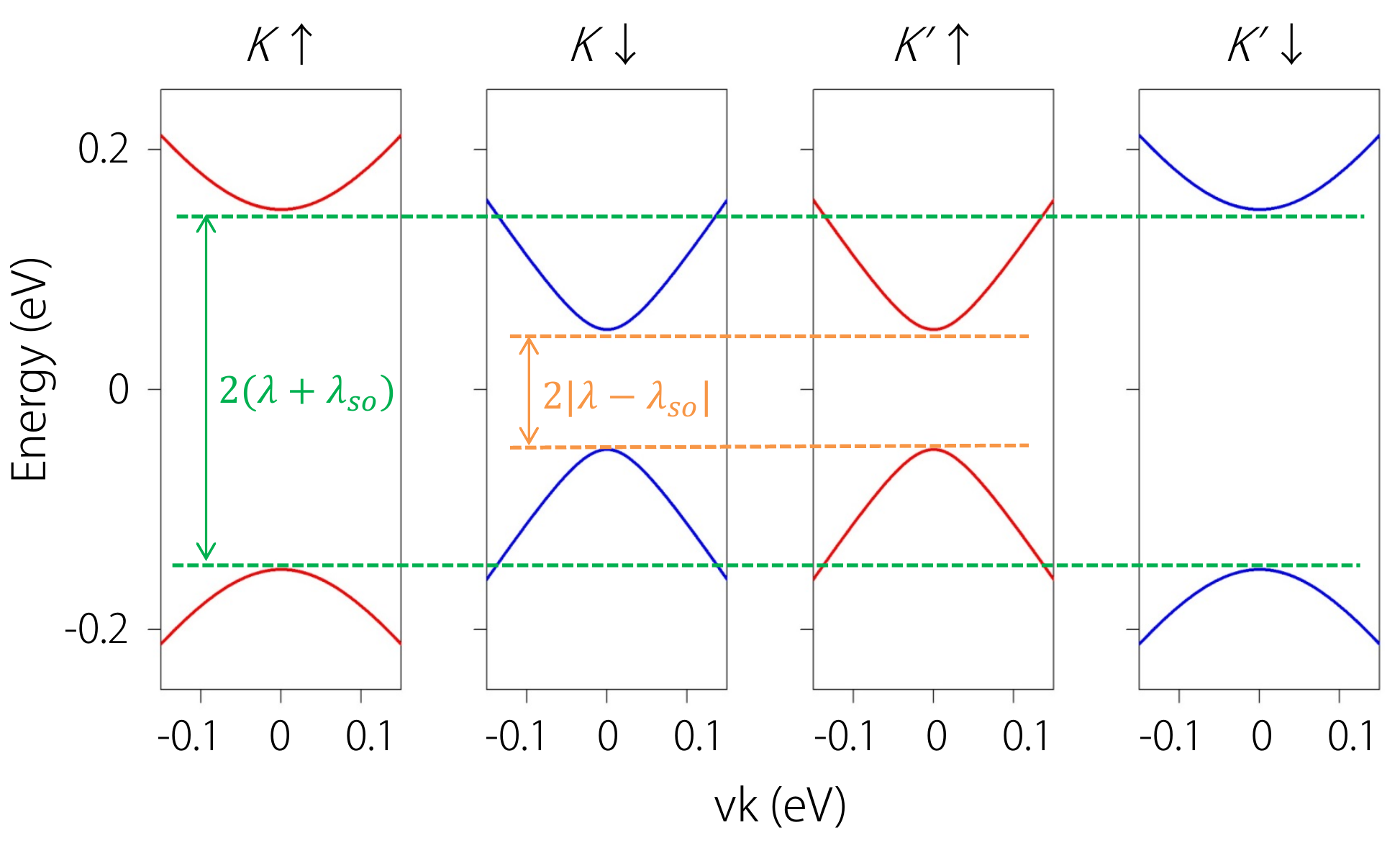}}
\caption{(color online) The energy dispersion for each spin-valley. Spin up (down) is marked with red (blue) color. The gap is $2(\lambda+\lambda_{so})$ for the upper bands and $2|\lambda-\lambda_{so}|$ for the lower bands. Assume both $\lambda$ and $\lambda_{so}$ are positive, the band gap is inverted when $\lambda_{so}>\lambda$. We have chosen the parameter values $v=0.5\times 10^6$~m/s, $\lambda=0.1$~eV, and $\lambda_{so}=0.05$~eV.}
\label{fig2}
\end{figure}

In this model, $s_z=\pm 1$ is a good quantum number, because the buckling in the considered materials is small and the mirror-plane symmetry breaking is weak. The model can thus be written as
\begin{equation}\label{HH}
H=\hbar v (\tau_z k_x\sigma_x+k_y\sigma_y)+\Delta_{\tau_z s_z}\sigma_z,
\end{equation}
with a combined gap parameter
\begin{equation}
\Delta_{\tau_z s_z}\equiv \lambda+\lambda_{so}\tau_z s_z,
\end{equation}
which depends on the product of valley and spin indices $\tau_z s_z=\pm 1$. Thus the Hamiltonian (\ref{HH}) reduces to two sets of massive Dirac fermions with flavor dependent mass terms $\Delta_{\tau_z s_z=\pm 1}$. In this paper, we will assume that both $\lambda$ and $\lambda_{so}$ are positive, and that the strength of $\lambda$ is controlled by an external electric field whereas that of $\lambda_{so}$ is intrinsic. When $\lambda=0$, at each valley the band is two-fold spin degenerate. As $\lambda$ increases from zero, the spin degeneracy is lifted, as shown in Fig.~\ref{fig2}. For $\lambda<\lambda_{so}$, the SOC gap dominates in the combined gap and the system is in a QSH insulator state. In the opposite case, when $\lambda>\lambda_{so}$ the system becomes topologically trivial. Evidently, the topological quantum phase transition occurs at the critical point $\lambda=\lambda_{so}$ where the gap $\Delta_-$ vanishes whereas $\Delta_+$ is enhanced.

Each flavor of $\tau_z s_z$ ($1$ or $-1$) at different valleys corresponds to opposite spins, i.e., a spin-valley locking. Moreover, the chirality (relaxed due to the energy gap) for the same flavor also differs between the two valleys. This can be easily observed by tracking how the pseudospin's in-plane component rotates around a constant energy surface at each valley.
These properties opposite at the two valleys will be of importance for the interesting physics discussed below.

\section{Landau Level Structure}
\begin{figure}[t]
\scalebox{0.44}{\includegraphics*{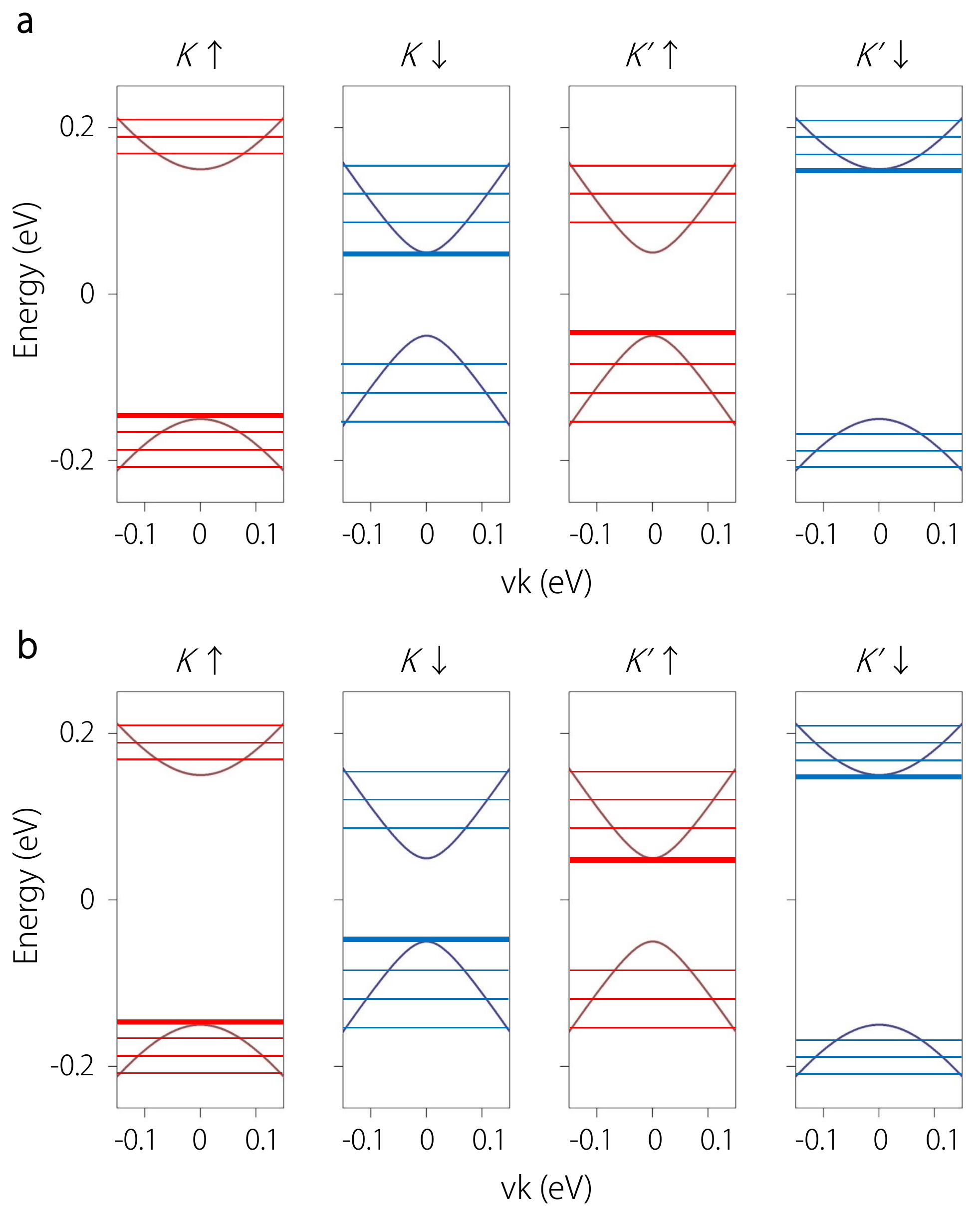}}
\caption{(color online) The first few LLs of each spin-valley resolved band for (a) the QSH insulator phase ($\lambda_{so}>\lambda>0$) and for (b) the trivial insulator phase ($\lambda>\lambda_{so}>0$).  The red (blue) color represents the spin up (down), and the $n=0$ LLs are marked with thicker lines with an asymmetric feature. Note that the positions of the $n=0$ LLs for the two lower bands ($\tau_z s_z=-1$) differ between the two phases. We have used the same parameter values as in Fig.~\ref{fig2}.}
\label{fig3}
\end{figure}

In the presence of a uniform perpendicular magnetic field $B$, the two-dimensional kinetic momentum $\hbar\bm k$ in Eq.~(\ref{HH}) is replaced by $\bm \pi=\hbar\bm k+e\bm A/c$, based on the standard Peierls substitution. In the Landau gauge the vector potential $\bm A$ takes the form of $\bm A=(0,Bx)$. Then we can define the bosonic ladder operators as $b^\dagger=(\ell_B/\sqrt{2}\hbar)\pi^+$ and $b=(\ell_B/\sqrt{2}\hbar)\pi^-$, where $\pi^\pm=\pi_x\pm i\pi_y$ and the magnetic length $\ell_B=\sqrt{\hbar/(eB)}=25.6/\sqrt{B[T]}$~nm. This model and all the following results are approximately valid when $\hbar v/\ell_B$ is smaller than the bandwidth of the effective Dirac model. The ladder operators satisfy the relations $[b,b^\dagger]=1$, $b|n\rangle=\sqrt{n}|n-1\rangle$, and $b|0\rangle=0$, where $|n\rangle$ is the $n$th LL eigenstate of a conventional 2D electron gas. Written in terms of the ladder operators and in the basis of $|n\rangle$, Hamiltonian~(\ref{HH}) can be easily diagonalized and the resulting spectrum reads
\begin{equation}\label{LL}
E_{n,\pm}=-\tau_z\Delta_{\tau_z s_z}\delta_{n,0}\pm\sqrt{\Delta_{\tau_z s_z}^2+n\hbar^2\omega_c^2}(1-\delta_{n,0})\,,
\end{equation}
where $\omega_c=\sqrt{2}v/\ell_B$ is the cyclotron frequency, $\delta$ is the Kronecker delta function, and $n$ is a non-negative integer denoting the LL orbitals. This spin-valley resolved LL structure is schematically shown in Fig.~\ref{fig3}. In the absence of the mass terms, e.g., in the case of graphene, the LLs are fourfold degenerate at each energy. For the case $\Delta_{\tau_z s_z}\neq 0$ the $SU(4)$ symmetry of the zeroth (or $n=0$) LLs are completely broken at the single-particle level, whereas all other LLs are broken into two groups with $\tau_z s_z=\pm 1$ and are doubly degenerate at each energy.

In particular, the $n=0$ LL energies $-\lambda\tau_z-\lambda_{so}s_z$ are independent of the magnetic field strength $B$. Evidently, the $SU(4)$ symmetry in the zero-mass case is broken between the two valleys as well as between the two spins. On one hand, when $\lambda_{so}>\lambda>0$, the two $n=0$ LLs of spin up are at the valence band top whereas the two of spin down are at the conduction band bottom, independent of their valley indices. In this scenario, the $\nu=0$ state has a quantized spin Hall conductivity that survives at $B=0$, reflecting the QSH state nature in the presence of an approximate mirror-plane symmetry.\cite{note-mirror}
On the other hand, when $\lambda>\lambda_{so}>0$, the two $n=0$ LLs of valley K are at the valence band top whereas the two of valley K' are at the conduction band bottom, independent of their spins. This scenario is consistent with the fact that the half filled $\nu=0$ state is adiabatically connected to the trivial insulating state at $B=0$, in which both the charge and spin Hall conductivities are zero.
The transition between the two scenarios occurs when $\lambda=\lambda_{so}$,
companied by a gap closure at two of the four spin-valleys with $\tau_z s_z=-1$.
The wavefunctions of $n=0$ LLs at valley K and K' are $(0,|0\rangle)^T$ and $(|0\rangle,0)^T$, respectively. Thus, for the $n=0$ LLs the valley and sublattice degrees of freedom coincide. This feature allows to tune the $n=0$ LL energies via the buckling of the two sublattices and the perpendicular electric field, namely, $\lambda_{so}$ and $\lambda$.
Note that we have neglected the roles of electron-electron interactions and Zeeman couplings to the electron spins and the atomic orbitals, and we will comment on these effects in Sec.~VIII.

We mention by passing that the asymmetric LL structure is a generic feature of massive Dirac fermions and is related to the opposite chirality of the two valleys and to the spin-valley dependent mass terms.\cite{xli2013,cai2013} One intuitive picture, as noted above, is to make a connection between the charge, spin, and valley Hall conductivities of the $\nu=0$ quantum Hall state and the classification of the $B=0$ states.\cite{fz2011} A more intuitive picture can be provided by the semiclassical theory of electron dynamics at low fields.\cite{cai2013} Due to the pseudospin-orbit coupling, a wave packet near a valley center has a self-rotation, which produces an intrinsic orbital magnetic moment\cite{fz2011,xiao2010}
\begin{equation}\label{littlem}
\bm m(\bm k)=-\tau_z\frac{ v^2m_e\Delta_{\tau_z s_z}}{(\Delta_{\tau_z s_z}^2+\hbar^2 v^2 k^2)}\mu_B\hat{\bm z}\,,
\end{equation}
where $m_e$ is the electron mass and $\mu_B=e\hbar/2m_e$ is the Bohr magneton. Note that the orbital moment is the same for the conduction and valence bands. In Eq.~(\ref{littlem}) the factor $\tau_z$ results from the opposite chirality of the two valleys whereas the factor $\Delta_{\tau_z s_z}$ reflects the role of the spin-valley dependent mass terms. The orbital moment couples with the perpendicular magnetic field $B$, shifts the LLs in an asymmetric way that is determined by the factor $\tau_z\Delta_{\tau_z s_z}$, and leads to the spin-valley resolved LL structure sketched in Fig.~\ref{fig3}.

\section{Landau Level Crossing Effect}
In the buckled honeycomb lattice, we can denote the group of LLs corresponding to flavor $\tau_z s_z=1~(-1)$ as group I (II).
At high fields, for large LL orbitals, and with small band gaps, $nB\gg\Delta_{\pm}^2/(2e\hbar v^2)$, the LL energies goes linearly with $\sqrt{B}$ as for the case of massless Dirac fermions. On the opposite limit, at low fields, for small LL orbitals, and with large band gaps, $nB\ll\Delta_{\pm}^2/(2e\hbar v^2)$, the LL energies goes linearly with $B$ as for the case in conventional quantum wells.
In the latter case, we can expand Eq.~(\ref{LL}) and write LL energies of group I as
\begin{equation}
E_{n_{\rm I},\pm}\simeq\pm\Delta_+\left(1+\frac{e\hbar v^2 n_{\rm I}B}{\Delta_+^2}\right)\,,
\end{equation}
where $n_\text{I}=0,1,2,\cdots$ labels the LLs of group I. By replacing $\Delta_+$ by $\Delta_-$ and $n_\text{I}$ by $n_\text{II}$ above, we obtain the corresponding expression for group II.

In general, when there is more than one channel of 2D conduction electrons,
their differences in velocity and in mass give rise to the LL crossing effect.
Such effects occur in the conventional quantum wells as a result of the Zeeman splitting between the spin up and spin down carriers,\cite{knob2002}
on the $(111)$ surface of SnTe due to the presence of the symmetry-unrelated $\bar{\Gamma}$ and $\bar M$ Dirac surface states,\cite{li2014}
in ABA trilayer graphene because of the chiral decomposition of
the monolayer-like and bilayer-like subbands,\cite{tayc2011,koshino2011,yuan2011,macd2012}
and in monolayer MoS$_2$ owing to the peculiar SOC of $d$-electrons in the valence bands.\cite{xli2013}

\begin{figure}[t]
\scalebox{0.48}{\includegraphics*{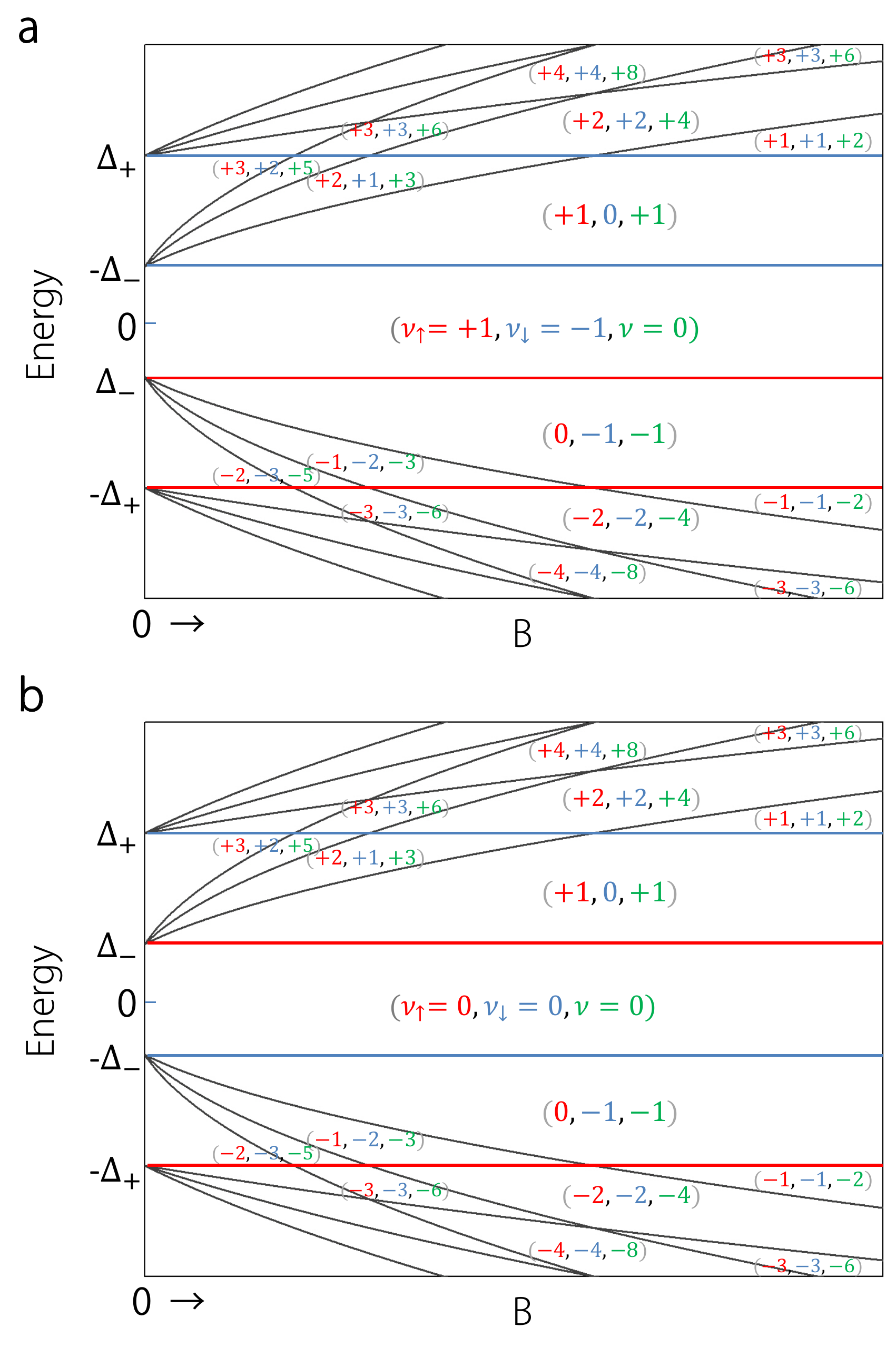}}
\caption{(color online) Schematic plot of the LL crossing pattern of the first few LLs for (a) the QSH insulator phase and for (b) the trivial insulator phase.  The $n=0$ LLs are spin-filtered while other levels are spin-degenerate. The red (blue) color represents spin up (down). In each gapped region, $(\nu_\uparrow,\nu_\downarrow,\nu)$ labels the filling factors of spin-up, spin-down, and total LLs.}
\label{fig4}
\end{figure}

For the case of buckled honeycomb lattice, the LL crossing effect must occur, since the two groups of LLs have different masses.
By equating the LL energies of the two groups in Eq.~(\ref{LL}), we find that the crossing point for two LLs with index $n_\text{I}$ and $n_\text{II}$ occurs at
\begin{eqnarray}
B_c=\frac{2\lambda\lambda_{so}}{e\hbar v^2(n_{\rm II}-n_{\rm I})}\,.
\end{eqnarray}
This result applies to both the conduction and the valence bands, since they are symmetric with respect to the zero energy, as shown in Fig.~\ref{fig4}.
Note that in those cases for $n_{\rm II},n_{\rm I}>0$ the crossing points are all fourfold degenerate,
whereas in those cases for $n_{\rm I}=0$ and $n_{\rm II}>0$ the crossing points are all threefold degenerate.
This is because both $n_{\rm I}=0$ LLs are non degenerate whereas all $n\neq0$ LLs are doubly degenerate.
The scenarios for both the QSH phase and the trivial phase are sketched in Fig.~\ref{fig4}.
Notably, the two scenarios only qualitatively differ in the $\nu=0$ cases.

When the electron-electron interactions are not substantial, as assumed in this paper,
the LL crossing effect further leads to the enhancement of longitudinal magnetoresistance in transport.
When the interactions become substantial in the presence of strong magnetic fields and weak disorders, small gaps may open at the crossing points and the magnetoresistance peaks split. More interestingly, the LL crossing effect disappears at $\lambda=0$, which can be tuned by an external electric field. Thus, the magnetoresistance in buckled honeycomb lattice can also be controlled by the electric field.

Even in the absence of interactions, the Hall plateaus follow an unconventional sequence:
$\nu=\cdots,-2M-4,-2M-2,-2M,-2M+1,\cdots,-3,-1,0,1,3,\cdots,2M-1,2M,2M+2,2M+4,\cdots$
Here the $n_{\rm I}=0$ LL lies between the LLs with $n_{\rm II}=M-1$ and $n_{\rm II}=M$, with $M$ given by
\begin{equation}
M=\left\lfloor \frac{2\lambda\lambda_{so}}{e\hbar v^2 B}\right\rfloor +1\,,
\end{equation}
where $\lfloor\cdots\rfloor$ is the floor function. The step of two in the sequence is a consequence of the $\tau_z s_z=\pm 1$ classification and the double degeneracy in each group of LLs with $n\neq 0$, a hallmark of the intrinsic SOC.
A step-one jump reflects the filling or emptying of a spin-filtered and valley-resolved $n=0$ LL. This is a direct result of broken $SU(4)$ spin-valley symmetry among the anomalous $n=0$ LLs, given by the masses $\Delta_{\pm}$ of the Dirac fermions.

\section{Fractionally Quantized Conductances in PN Junctions}
The unique band structure of buckled honeycomb-lattice materials allows reconfigurable electric-field control of carrier type and density, making them ideal candidates for bipolar nanoelectronics. A p-p, n-n, or p-n junction can be realized by using electrostatic gating to independently control the local carrier types and densities in two adjacent regions.\cite{will2007,ozyi2007} In such a device made of a buckled honeycomb-lattice material, transport measurements in the quantum Hall regime can reveal new plateaus with integer and fractionally quantized two-terminal conductances across the junction. This effect arises from the redistribution of quantum Hall currents among edge channels propagating along and across the junction, due to the presence of residue disorder.\cite{xli2013,aban2007,carm2010,carm2011}

In the presence of magnetic disorders or strong mirror symmetry breaking, $s_z$ is not conserved. Because of the redistribution of the chiral quantum Hall edge currents at the junction, the net conductance in units of $e^2/h$ across the junction is quantized as
\begin{eqnarray}\label{Gmix1}
G_{pp,nn}&=&\mbox{min}\{|\nu_1|,|\nu_2|\}\,,\\
\label{Gmix2}
G_{pn}&=&\frac{|\nu_1||\nu_2|}{|\nu_1|+|\nu_2|}\,,
\end{eqnarray}
in the unipolar and bipolar regimes, respectively. Here $\nu_1$ and $\nu_2$ are the filling factors of the two sides of the junction. This limit is similar to the case of graphene where all possible filling factors are even. Consider the special case with $\nu_1=0$ and $\nu_2=2n$, it follows that $G=0$. We note that in this case the conductance is insensitive to whether the $B=0$ phase is topological or trivial.

When the residue disorder is nonmagnetic and the mirror symmetry is approximately preserved,
$s_z$ can be considered as a good quantum number and the $n=0$ LLs are spin filtered.
It follows that the full equilibrium must be achieved within each $s_z$ subspace independently.
Consequently, the net conductance across the junction is
\begin{eqnarray}\label{Gnmix}
G=\sum_{s_z=\uparrow,\downarrow}\bigg[\mbox{min}\{|\nu_{1s_z}|,|\nu_{2s_z}|\}\Theta(\nu_{1s_z}\nu_{2s_z})\nonumber\\
+\frac{|\nu_{1s_z}||\nu_{2s_z}|}{|\nu_{1s_z}|+|\nu_{2s_z}|}\Theta(-\nu_{1s_z}\nu_{2s_z})\bigg]\,,\label{pn-s}
\end{eqnarray}
where $\Theta$ is the Heaviside step function, the first and the second terms are the spin-resolved conductances in the unipolar and the bipolar regimes, respectively.
The total filling factors are implied by $\nu_1=\nu_{1\uparrow}+\nu_{1\downarrow}$ and $\nu_2=\nu_{2\uparrow}+\nu_{2\downarrow}$.
As we will see shortly, in the QSH phase the two spin species are not necessarily in the same bipolar or unipolar regime,
whereas in the trivial phase both spin species are always in the same regime. Such inconsistency is quite counterintuitive, reflecting the topological quantum phase transition at $\lambda=\lambda_{so}$ for the $B=0$ case.

Only when $\nu_{\uparrow}\neq\nu_{\downarrow}$ on at least one side of the junction, the conductance in Eq.~(\ref{pn-s}) is essentially different with the simple case (with magnetic disorders) in which only the total filling factors $\nu_1$ and $\nu_2$ matter. In the absence of an electric field, i.e., $\lambda=0$, consider a junction with $\nu_1=0$ and $\nu_2=4n-2$ for some integer $n$,  This indicates that $\nu_{1\uparrow}=-\nu_{1\downarrow}=1$ and $\nu_{2\uparrow}=\nu_{2\downarrow}=2n-1$. From Eq.~(\ref{pn-s}) we find that $G=(4n-1)/(2n)$ for $n>0$ and that $G=(3-4n)/(2-2n)$ for $n\leq0$.

\begin{figure}[t]
\scalebox{0.50}{\includegraphics*{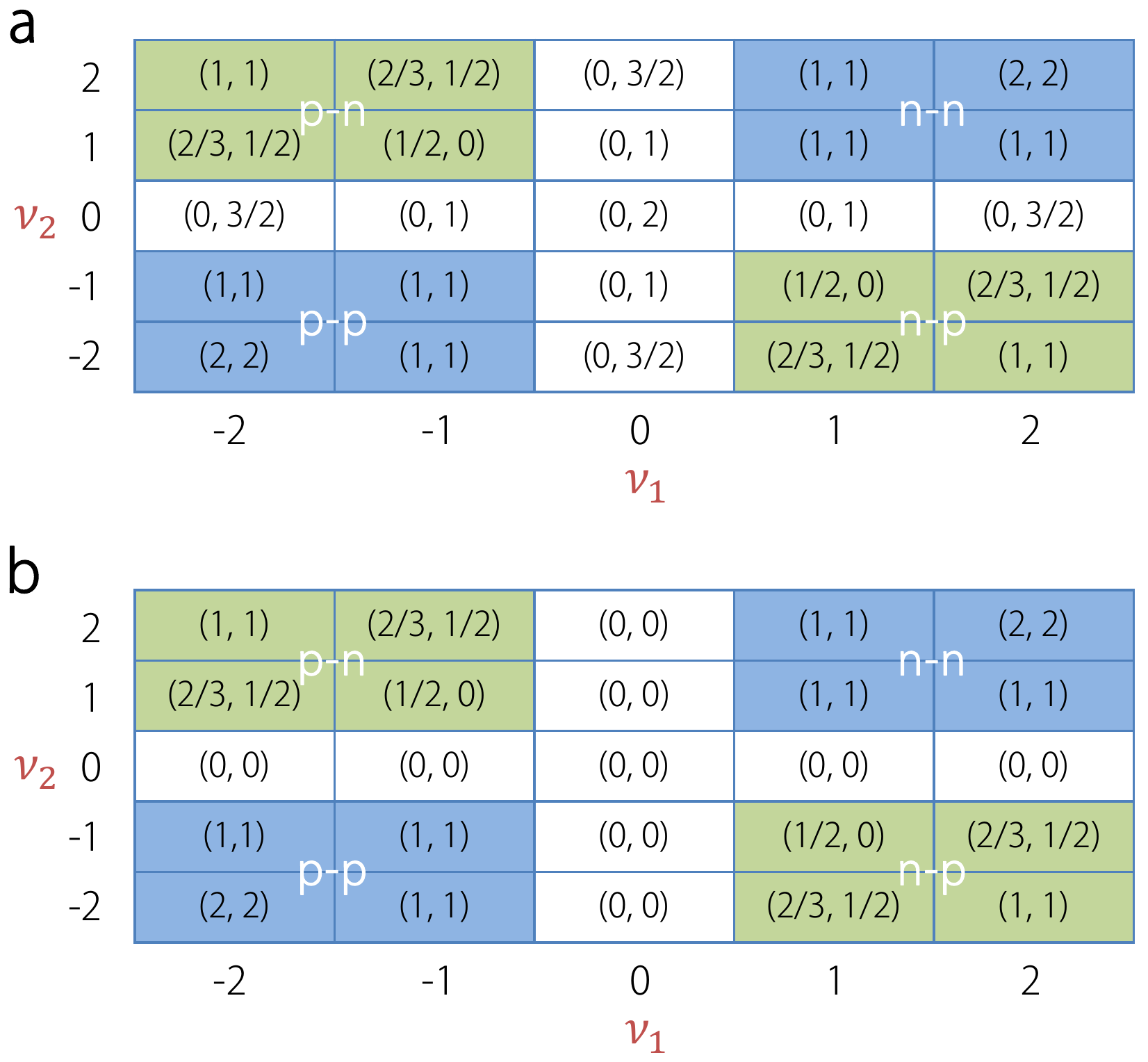}}
\caption{(color online) Map of the conductance in units of $e^2/h$ across the p-p, n-n, and p-n junctions, as a function of $\nu_1$ and $\nu_2$ ($|\nu_1|,|\nu_2|\leq 2)$. (a) is for the QSH phase and (b) is for the trivial phase. In each entry $(G_\alpha, G_\beta)$, $G_\alpha$ is the conductance with spin mixing whereas $G_\beta$ is the conductance without spin mixing.}
\label{fig5}
\end{figure}

We then consider the scenario of $\lambda_{so}>\lambda>0$, in the presence of a small perpendicular electric field. The conductance is unchanged for the above case with $\nu_1=0$ and $\nu_2=4n-2$. Now imagine while fixing $\nu_1=0$ we tune the gate in region two such that $\nu_2=2m-1$ for some integer $m$. The latter filling indicates that $\nu_{2\uparrow}=\nu_{2\downarrow}+1=m$. Thus, $G=(2m-1)/m$ for $m>0$ and $G=(1-2m)/(1-m)$ for $m\leq0$.

When the electric field is sufficiently large to invert the band gap such that $\lambda>\lambda_{so}>0$, in this scenario, $\nu_1=0$ would indicate that $\nu_{1\uparrow}=\nu_{1\downarrow}=0$. Thus, $G=0$ as long as one region of the junction is half filled, in sharp contrast to the QSH phase.

Although we have focused on the case with $\nu_1=0$ to illustrate the essence of the physics, we exhaustively show the full map of conductance in Fig.~\ref{fig5} for $|\nu_1|,|\nu_2|\leq 2$. The two numbers in a parenthesis are the net conductances with and without spin mixing respectively. Fig.~\ref{fig5}(a) is for the QSH phase whereas Fig.~\ref{fig5}(b) is for the trivial phase. One observes that the difference between the two phases is exhibited for the cases with either $\nu_1=0$ or $\nu_2=0$, as we discussed before. One may also consider the cases in which the two sides of the junction are in different phases. The analysis is straightforward and our results Eqs.~(\ref{Gmix1}), (\ref{Gmix2}) and (\ref{Gnmix}) still apply. These above mentioned interesting features are indeed richer than those in graphene\cite{will2007,ozyi2007,aban2007,carm2010,carm2011} and transition metal dichalcogenides.\cite{xli2013}
Thus, the conductance across the junction, distinct in the two phases, can serve as a useful diagnosis for the phase of buckled honeycomb-lattice material under an electric field.

In addition to the unconventional transport properties, STM probes at the interface can also detect a special fingerprint of the spin-filtered $n=0$ LLs. This is the case as long as the Fermi energies of the two regions lie in different energy windows that are divided by the $n=0$ LLs (see Fig.~\ref{fig3}).
In particular, spin-filtered edge states, whose number is given by $|\nu_{1s_z}-\nu_{2s_z}|$, will propagate along the interface. The interface current can be controlled in the following senses. (i) Switching the magnetic field direction flips the spin polarization of the current. (ii) Interchanging $\nu_1$ and $\nu_2$ switches the current direction while tuning $\nu_1$ and $\nu_2$ adjusts the current amplitude. (iii) Tuning one Fermi energy to a different energy window while fixing the other one may change the carrier type, besides the effects in (i) and (ii). (iv) Most importantly, as we have analyzed above, the electric field can tune the integer or fractionally quantized conductance in an unprecedented way,
and even diagnose the topological nature of the phase of the material.

\section{Electric-Field Control of Spin, Valley, and Sublattice Polarizations}
In order to use the valley degrees of freedom for information processing, it is necessary to have efficient ways to generate and control the valley polarization of carriers. This is similar to the case of spintronics, where the task of generating spin polarizations has been a topic of active research in the past decades. While spin can be easily coupled to Zeeman fields and controlled through spin-orbit couplings, the manipulation of valley degrees of freedom is much more challenging. Recently, based on an argument of orbital magnetic moment, it has been demonstrated that valley polarization could be generated by a circularly polarized light in transition metal dichalcogenides.\cite{zeng2012,mak2012,cao2012}
Furthermore, it has been proposed that a magnetic orbital field can be used to control the valley polarization in transition metal dichalcogenides.\cite{cai2013}
A similar effect is also present and even constitute two advances in the buckled honeycomb-lattice materials.
First, the band gap and the topological nature of the phase can be tuned by an external electric field,
providing more freedom in controlling the valley degrees of freedom.
Secondly, for both electron-doped and hole-doped cases, because the splitting between $\tau_z s_z=1$ and $-1$ bands lead to the spin-valley locking, the valley polarization amounts to a spin polarization, allowing to manipulate valley degrees of freedom by means of spintronics.
Here we focus on the electric-field control of the spin, valley, and sublattice polarizations in the strong magnetic field regime.

We start from the case in which the electric field is zero, $\lambda=0$. From the LL structure Eq.~(\ref{LL}), at a fixed chemical potential, there is no valley polarization, but there exists finite spin polarization of the charge carriers. This is easily understood by noticing that the four $n=0$ LLs are valley degenerate but spin split. For example, the $n=0$ LLs at the conduction (valence) band bottom (top) for the two valleys are both of spin down (up). All higher LLs are spin and valley degenerate. As a result, the valley and spin polarizations are respectively given by
\begin{eqnarray}
P_v&\equiv& \nu_+-\nu_-=0\,,\\
P_s&\equiv& \nu_\uparrow-\nu_\downarrow=2\delta_{\nu,0}\,,
\end{eqnarray}
where $\nu_\pm$ and $\nu_{\uparrow,\downarrow}$ are the partial filling factors for valley K and K' and spin up and down, respectively.
Note that $P_v$ and $P_s$ are simply the quantized valley and spin Hall conductivities in units of $e^2/h$.
For the present case with $\lambda=0$, it is the valley degeneracy that dictates $P_v$ to be zero and $\nu$ to take the form of $4(n+1/2)$, with $n$ being an arbitrary integer.
The spin polarization is maximized at $\nu=0$, i.e., when the two $n=0$ LLs of spin $\uparrow$ ($\downarrow$) are occupied (empty),
independent of the strength of magnetic field. The spin polarization disappears when the $n=0$ LLs become completely occupied or empty.

When a small electric field is applied such that $\lambda_{so}>\lambda>0$, the LLs split into two groups with $\tau_z s_z=\pm 1$.
As shown in Fig.~\ref{fig3}(a), all the LLs are doubly degenerate except the four non-degenerate $n=0$ LLs; in ascending order of energy,
these four LLs are indexed by spin up and valley K, spin up and valley K', spin down and valley K, and spin down and valley K'.
When one or three $n=0$ LLs are filled, the filling factor $\nu$ is odd, and both the spin and the valley polarizations are one.
When two of them are filled, $\nu$ becomes zero and the spin polarization is maximized whereas the valley polarization vanishes.
When all of them are filled or empty, $\nu$ is even and both polarizations are zero. Therefore,
\begin{eqnarray}
P_v&=&\delta_{\nu,2n-1}\,,\\
P_s&=&\delta_{\nu,2n-1}+2\delta_{\nu,0}\,
\end{eqnarray}
where $n$ is an arbitrary integer.

Further increasing the electric field such that $\lambda>\lambda_{so}>0$, the two middle $n=0$ LLs switch their energy orders, as seen in Fig.~\ref{fig3}(b). This follows from the topological quantum phase transition between the QSH and quantum valley Hall phases at $B=0$. (The latter phase is also referred as a trivial phase in other sections.)
As an interesting result, spin and valley switch their roles. Therefore, we can anticipate that
\begin{eqnarray}
P_s&=&\delta_{\nu,2n-1}\,,\\
P_v&=&\delta_{\nu,2n-1}+2\delta_{\nu,0}\,
\end{eqnarray}
where $n$ is an arbitrary integer again.

The above results of spin and valley polarizations for states with filling factors $|\nu|\leq 3$ are listed in Table~\ref{tableone}. One notes that the coupled valley and spin polarizations are non-vanishing at odd filling factors. Here, the difference between the QSH phase and the trivial phase is reflected in the $\nu=0$ case: for QSH phase, it is spin polarized but valley non-polarized, whereas the situation is reversed for the trivial phase.

\begin{table}[t!]
\caption{Spin and valley polarizations $(P_s, P_v)$ as functions of the filling factor $\nu$
for both the QSH phase ($\lambda_{so}>\lambda>0$) and the trivial phase
($\lambda>\lambda_{so}>0$). }
\newcommand\T{\rule{0pt}{3.0ex}}
\newcommand\B{\rule[-1.7ex]{0pt}{0pt}}
\centering
\begin{tabular}{c|ccccccc}
      \hline\hline
      filling factor $\nu$ & -3 & -2 & -1 & 0 & 1 & 2 & 3\\[1pt]
      \hline\\
      QSH $(P_s, P_v)$ & (1,1) & (0,0) & (1,1) & (2,0) & (1,1) & (0,0) & (1,1)\\[1pt]\\
      Trivial $(P_s, P_v)$ & (1,1) & (0,0) & (1,1) & (0,2) & (1,1) & (0,0) & (1,1)\\[1pt]\\
      \hline\hline
\end{tabular}
\label{tableone}
\end{table}

The simultaneous polarization of carriers in both valley and spin permits versatile methods for their detection and manipulation.
Moreover, the polarization reversal occurs at the transition between the topological phase and the trivial phase also offers a way to experimentally differentiate them.

To close this section, we note by passing that for $n=0$ LLs the valley pseudospin coincides with the sublattice pseudospin, as the $n=0$ LLs of a particular valley completely localize on a particular sublattice. Thus, the discussed valley polarization, induced by the peculiar $n=0$ LLs, is equivalent to the sublattice polarization.

\section{Anomalous Hall Transport}
The transverse motion of carriers can be induced even in the absence of external magnetic field, by the Berry curvature of the electronic band structure. The Berry curvature acts like a magnetic field in the reciprocal space, which leads to a transverse velocity term in the semiclassical equation of motion for electron wave packets.\cite{chang1996,sund1999} In the presence of sublattice (chiral) symmetry, the Berry curvature is required~\cite{TMSC} to vanish for a gapped system, otherwise the system must be gapless.
The massive Dirac fermions in a buckled honeycomb lattice acquire a finite Berry curvature,
since the sublattice symmetry is explicitly broken by the SOC and by the electric field.
For our model~(\ref{Ht}), the Berry curvature is given by
\begin{equation}\label{BerryC}
\bm{\mathit{\Omega}}(\bm k)=-\alpha\tau_z\frac{\hbar^2 v^2\Delta_{\tau_z s_z}}{2(\Delta_{\tau_z s_z}^2+\hbar^2 v^2 k^2)^{3/2}}\hat{z}.
\end{equation}
where $\alpha=\pm$ denotes the conduction and valence bands.
Similar to the orbital magnetic moment, in Eq.~(\ref{BerryC}) the factor $\tau_z$ results from the opposite chirality of the two valleys whereas the factor $\Delta_{\tau_z s_z}$ reflects the role of the spin-valley dependent mass terms that break the sublattice symmetry.
The integral of Berry curvature over all the filled states gives the intrinsic contribution to the Hall conductivity:\cite{jung2002,onod2002}
\begin{equation}\label{intH}
\sigma_H=\frac{e^2}{h}\sum_{n} \int\frac{d^2k}{(2\pi)^2}f_n(\bm k)\mathit\Omega_n(\bm k)\,,
\end{equation}
where $f_n(\bm k)$ is the Fermi distribution at state $|n,{\bm k}\rangle$ and $n$ is a band index.

First, in the absence of external electric fields, $\lambda=0$ and $\Delta_+=-\Delta_-=\lambda_{so}$.
When the Fermi level is in the band gap, the system is a QSH insulator, with
\begin{equation}\label{qsh}
\sigma_{H}^\uparrow=-\sigma_{H}^\downarrow=\frac{e^2}{h}\,,\qquad \sigma_H^s=\frac{e^2}{h}\,,
\end{equation}
where $\sigma_{H}^{\uparrow(\downarrow)}$ is the Hall conductivity for spin up (down), and $\sigma_H^s=(\sigma_{H}^\uparrow-\sigma_{H}^\downarrow)/2$ is the spin Hall conductivity. This quantized spin Hall conductivity is related to the spin helical edge states for a finite system.
Here we emphasize that the $s_z$ conservation follows from the assumption that the mirror symmetry is only weakly broken.
Furthermore, different from the longitudinal transport, the anomalous Hall transport involves contribution from all the occupied states, as implied in Eq.~(\ref{intH}), and thus must go beyond the low energy model.
However, the Hall conductivity must be quantized in unit of $e^2/h$ when the Fermi level lies in a gap. Hence the results in Eq.~(\ref{qsh}) should be understood as valid up to an integer multiple of $e^2/h$, and this integer must be even for a $Z_2$ QSH insulator.
Nevertheless, the determination of this integer would require the knowledge of the full band structure.
These two points also apply to the following discussions.

We are more interested in the case with finite doping, in which the Hall conductivity is not quantized. We shall mainly discuss the $n$-doped case. The results for $p$-doped case can be easily obtained by a similar procedure. In the metallic case, the Hall conductivity has additional contributions from scattering of carriers around the Fermi energy.\cite{sini2008,naga2010} There is an important side jump contribution\cite{berg1970} that is proportional to the Berry curvature at the Fermi energy. Here we shall take a simple Gaussian white-noise scattering model\cite{gaussian} and disregard the intervalley scattering which requires a large momentum transfer. For each flavor the Hall conductivity including both intrinsic and extrinsic contributions is given by
\begin{equation}\label{sigst}
\sigma_H^{s_z,\tau_z}=-s_z\frac{e^2}{2h}\left[1-\frac{\lambda_{so}}{\mu}-\frac{\lambda_{so}(\mu^2-\lambda^2_{so})}{\mu^3}\right]\,,
\end{equation}
where $\mu$ ($\mu>\lambda_{so}$) is the chemical potential. This result only take into account the contribution from conduction band carriers. From this, we easily obtain the spin and valley Hall conductivities
\begin{equation}
\sigma^s_H=\frac{e^2}{h}\left[\frac{\lambda_{so}}{\mu}+\frac{\lambda_{so}(\mu^2-\lambda^2_{so})}{\mu^3}\right]\,,\qquad
\sigma^v_H=0\,.
\end{equation}
Note that the valley Hall conductivity defined as $\sigma_H^v\equiv (\sigma_H^{\tau_z=+1}-\sigma_H^{\tau_z=-1})/2$ vanishes because for each valley the contributions from the two spins cancel each other.

When a perpendicular electric field is applied, the bands for the two flavors $\tau_z s_z=\pm 1$ split. Consider the weakly doped case such that only the $\tau_z s_z=-1$ conduction bands are partially occupied, i.e., $|\Delta_-|<\mu<\Delta_+$. For $0<\lambda<\lambda_{so}$, similar to Eq.~(\ref{sigst}), we find that
\begin{equation}\label{sigsz}
\sigma_H^{s_z}=-s_z\frac{e^2}{2h}\left[1-\frac{|\Delta_-|}{\mu}-\frac{|\Delta_-|(\mu^2-\Delta_-^2)}{\mu^3}\right]\,,
\end{equation}
which leads to the spin and valley Hall conductivities
\begin{flalign}\label{sigs}
&\sigma^s_H=\frac{e^2}{2h}\left[\frac{|\Delta_-|}{\mu}+\frac{|\Delta_-|(\mu^2-\Delta^2_-)}{\mu^3}\right],\;
\sigma^v_H=-\sigma^s_H.
\end{flalign}
Note that there also exists a finite valley Hall conductivity in this case, owing to the spin-valley locking $\tau_z s_z=-1$.

When the gap is inverted by further increasing the electric field, i.e. $\lambda>\lambda_{so}$, the Berry curvatures for the two spin-valleys with $\tau_z s_z=\pm 1$ switch signs after the gap closes and reopens. As a result, both the spin Hall conductivity and the valley Hall conductivity in Eq.~(\ref{sigs}) change signs. Therefore, the sign change in the spin or valley Hall conductivity can be used to detect the topological quantum phase transition, induced by the electric field.

The above results are derived in the presence of time-reversal symmetry and the charge Hall conductivity must be zero.
The charge Hall conductivity becomes nonzero when the time-reversal symmetry is explicitly broken by an applied magnetic field.
In the high-field regime, the Hall conductivity becomes quantized due to the formation of LLs and follows an unconventional sequence,
as we discussed in Sec.~III and IV. Here, instead, we are concerned with the low-field regime where a semiclassical description is applicable.
In this picture, the effect of magnetic field is twofold. First, it exerts a Lorentz force on the carriers leading to an ordinary Hall effect.
Secondly, it couples with the orbital magnetic moment~(\ref{littlem}) and shifts the band energy as $-\bm m\cdot \bm B$.\cite{xiao2010,fz2011}
Because the moment has opposite signs between the two valleys, the relative energy shift between the two valleys gives rise to an anomalous contribution to the charge Hall effect. The ordinary Hall conductivity is well known as $\sigma_H^\text{ord}\simeq\rho_H^\text{ord}/\rho^2$, where $\rho_H^\text{ord}=-B/(en)$ is the ordinary Hall resistivity and $\rho$ is the longitudinal resistivity. In the following, we focus on the weakly doped case $|\Delta_-|<\mu<\Delta_+$, where only the $\tau_z s_z=-1$ conduction bands are partially occupied. (The inclusion of $\tau_z s_z=+1$ bands when $\mu>\Delta_+$ is straightforward and in fact decreases the anomalous effect.)

When $0<\lambda<\lambda_{so}$, the coupling $\delta E=-\bm m\cdot\bm B$ shifts the $\tau_z s_z=-1$ conduction band at valley K (K') down (up), according to Eq.~(\ref{littlem}). The relative shift between the band bottoms at two valleys is
\begin{equation}
\delta\mu=2mB\simeq \frac{e\hbar v^2}{|\Delta_-|}B.
\end{equation}
From Eq.~(\ref{sigsz}), we observe that the contributions to the charge Hall conductivity from the two valleys have opposite signs. The energy shift $\delta\mu$ breaks the perfect cancellation between them and leads to a net charge Hall contribution from the more populated valley
\begin{equation}\label{aH}
\delta\sigma_H^{c}\simeq \frac{e^2}{h}\frac{|\Delta_-|}{\mu^2}\left(1-\frac{3\Delta_-^2}{2\mu^2}\right)\delta\mu,
\end{equation}
where we have assumed that $\delta\mu\ll \mu$ for the low-field case. We note that when $\lambda>\lambda_{so}$, both the Berry curvature~(\ref{BerryC}) and the orbital magnetic moment~(\ref{littlem}) change signs, hence the anomalous contribution still has the form as in Eq.~(\ref{aH}). Furthermore, this anomalous charge Hall conductivity is proportional to the magnetic field strength. Thus, the corresponding Hall current $\delta j_H$ is a nonlinear response to the external electric field. We also note that this contribution could have a sign change at $\mu=\sqrt{3/2}|\Delta_-|$, which can be traced back to the different chemical potential dependence between the intrinsic and the side jump terms.
The ratio between the anomalous and the ordinary contributions is
\begin{equation}
\frac{\delta\sigma_H^{c}}{\sigma^\text{ord}_H}=-\left(1-\frac{\Delta_-^2}{\mu^2}\right)
\left(1-\frac{3\Delta_-^2}{2\mu^2}\right)\left(\frac{e^2}{h}\rho\right)^2.
\end{equation}
Evidently, this ratio depends on the resistivity in unit of $h/e^2$ and the anomalous part is more important for dirty samples with a large longitudinal resistivity.

\section{Discussion}

We have considered how the orbital magnetic moment, a pure lattice effect related to the Berry curvature, emerges in the Bloch bands and couples to the orbital magnetic fields.
However, we have neglected the roles of electron spins and atomic orbitals which naturally couple to the Zeeman fields.
As we will see, this treatment is indeed reasonable in those buckled honeycomb-lattice materials we have mentioned above.
Both spin and atomic orbital ($p$-orbitals here) related magnetic moments are about $\mu_B\sim e\hbar/(2m_e)$, whereas based on Eq.~(\ref{littlem}) the orbital magnetic moment at $k=0$ is
\begin{equation}
\mu_B^*=\frac{e\hbar v^2}{2\Delta}\,,
\end{equation}
where $v$ is the Fermi velocity, $\Delta$ the Dirac mass is half the gap size. Evidently, the Zeeman splitting is subdominant when
$\mu_B^*\gg \mu_B$, which requires
\begin{equation}
m_e v^2/\Delta\gg 1\,.
\end{equation}
For $m_e=0.51\times 10^6$~eV/c$^2$ and a typical value $v\sim 0.5\times 10^6$~m/s, the condition reduces to $\Delta\ll 1.4$~eV.
In general, $\Delta\ll 0.5$~eV holds for silicene, germanene, and most $X$-hydride/halide ($X$=N-Bi) monolayers. Therefore, we conclude that the Zeeman splitting is subdominant in these materials.

The above discussion is for the small magnetic field case. In relative larger fields, the Zeeman splitting should be much smaller than the LL gaps, although it further break the twofold degeneracy of the obtained LL structure. However, we do emphasize that the Coulomb exchange interaction, with an energy scale
\begin{equation}
\frac{e^2}{\epsilon\ell_B}=\frac{28}{\epsilon}\sqrt{B[T]}~\mbox{meV}
\end{equation}
should completely lift the spin degeneracies of LLs in the case of very strong magnetic fields and very weak disorders. This many-body interaction induced Zeeman effect is way larger than the single-particle Zeeman effects due to spins and atomic orbitals and may even turn over the energy order of the $n=0$ LLs given in the spectrum~(\ref{LL}). Nevertheless, we focus on the single-particle phenomena in this paper and leave the details of quantum Hall ferromagnetism to future works.

Finally, we point out that the buckling in some honeycomb-lattice materials lead to the emergence of an electron pocket at the $\Gamma$ point.\cite{xu2013,song2014} It is true that this extra valley adds complexity to the band structure and the corresponding LL spectrum. However, the $\Gamma$ pocket behaves like a conventional single-band 2D electron gas (2DEG) system. Thus, we expect that the main results would not be changed. For the LL structure that we are extremely interested in, the anomalous features of the $n=0$ LLs remain the same, although a conventional LL plateau sequence is superimposed over the unconventional sequence we find in Sec.~IV.

In summary, we have investigated the quantum and the anomalous Hall transport phenomena of a class of buckled honeycomb-lattice materials in response to an applied magnetic (orbital) field, with emphases on the tuning effect of an electric field.
Furthermore, in a p-n junction geometry we have explored some additional Hall plateaus for these materials, as ideal candidates for bipolar nanoelectronics.
Lastly, we have argued the roles of electron-electron interactions, the Zeeman couplings to electron spins and atomic orbitals, and the extra electron pocket at $\Gamma$ point.
Our study would facilitate the investigations on the 2D buckled honeycomb-lattice materials and help the design of novel electronic devices that may combine the charge, spin, valley, and sublattice degrees of freedom to achieve better performance and unprecedented functionalities.

{\indent{\em Acknowledgments.}}--- We would like to thank Chuanwei Zhang, Yugui Yao, and D. L. Deng for helpful discussions. S.A.Y. is supported by SUTD-SRG-EPD2013062. H.P. is supported by NSFC Grant No. 11174022. F.Z. is supported by UT Dallas research enhancement funds.

\bibliographystyle{apsrev4-1}

\begin{thebibliography}{100}
\bibitem{note-mirror} The time-reversal symmetry protects a $Z_2$ topological insulator,
whereas the mirror-plane symmetry protects a $(Z_\uparrow, Z_\downarrow)$ topological insulator.
When both symmetries are present, $Z_\uparrow + Z_\downarrow = 0$ and $Z_\uparrow~\mbox{mod}~2 = Z_2$.
\end{thebibliography}

\end{document}